\documentclass[aps,twocolumn,floatfix,showpacs]{revtex4}
\usepackage{graphicx}
\begin{document}

\title{A Uniform Approximation for the Coherent State Propagator
using a Conjugate Application of the Bargmann Representation}

\author{A. D. Ribeiro, M. Novaes and M. A. M. de Aguiar}

\affiliation{Instituto de F\'{\i}sica ``Gleb Wataghin'',
Universidade Estadual de Campinas, 13083-970, Campinas, S\~ao
Paulo, Brazil}

\begin{abstract}

We propose a conjugate application of the Bargmann representation
of quantum mechanics. Applying the Maslov method to the
semiclassical connection formula between the two representations,
we derive a uniform semiclassical approximation for the coherent
state propagator which is finite at phase space caustics.

\end{abstract}


\maketitle


Semiclassical approximations have been widely used in many areas
of physics. They are fundamental to the conceptual understanding
of the quantum-classical connection and are also very important in
practical situations where quantum calculation are difficult, as
in systems with many of degrees of freedom or with complicated
potential functions. However, semiclassical formulae are often not
globally valid, i.e., they are not appropriate to describe the
corresponding quantum function in all regions of the space of
parameters. The WKB formula for the eigenfunctions of a particle
in a one-dimensional potential well provides a simple example
\cite{berrymount}: on the classically allowed side of the well the
wave-function is oscillatory, whereas on the classically forbidden
side it is given by a single decreasing exponential. At the
boundary between the two regions, the turning point, the WKB
formula becomes singular. The formula actually fails in a whole
neighborhood of the singularity, whose size goes to zero as
$\hbar$ goes to zero. For non-stationary wavefunctions the
singularities occur at focal points or caustics. After a focal
point, but sufficiently away from it, the semiclassical formulae
for wave-functions or propagators still provide good
approximations, provided the proper Morse phases are added
\cite{berry83}.

This general problem of semiclassical expressions, which leads to
divergences and discontinuities in the semiclassical results, can
usually be eliminated by properly connecting the semiclassical
expressions on the different regions of validity and eliminating
spurious contributions. The most direct way to do that is to solve
the Schr\"{o}dinger equation in the vicinity of the singularity
and extend the solution towards the two regions. For the WKB
problem this amounts to linearize the potential about the turning
point, leading to the well known solution involving the Airy
function \cite{berrymount}. For non-stationary wavefunctions,
however, this approach is not usually possible and the Maslov
method has to be used \cite{berry83,maslov,maslov2}. It consists
basically in changing to a dual representation, where the
singularity does not exist. For a singularity in coordinates, one
uses the momentum representation and vice-versa. The trick is
that, when transforming back to the original representation, one
should go beyond the quadratic approximation, otherwise the
singularity re-appears. Usually, a stationary phase approximation
with the exponent expanded up to cubic terms is enough, giving
rise once again to corrections involving Airy functions.

In this Letter, we study the singularities of the semiclassical
propagator in the coherent state representation. These
singularities, called Phase Space Caustics (PSC), have been first
identified in \cite{adachi} and later studied in
\cite{Klau95,tanaka98,ribeiro1}. All these previous works were
concerned with ways to identify the singularities and prune the
branches of spurious contributions arising from them. Here we
tackle the problem of how to improve the semiclassical formula in
order to avoid its divergence at the caustics. This is a very
peculiar situation, since the phase space representation provided
by the coherent states makes use of both coordinate and momentum,
leaving no room for a natural dual representation. In this paper
we define an application that works as the canonical conjugate of
the Bargmann representation \cite{bargmann} and we use it to
derive a uniform semiclassical formula for the coherent state
propagator valid in the vicinity of the phase space caustics. For
the sake of clarity, we restrict ourselves to systems with one
degree of freedom. Results for multidimensional systems, which can
be treated along the same lines, and detailed numerical
applications will be published elsewhere.

The non-normalized coherent state $|z_0\rangle$ is defined as
\begin{equation}
|z_0\rangle =  e^{z_0 \hat{a}^{\dagger}}
   |0\rangle ,
\end{equation}
where
\begin{equation}
\hat{a}^{\dagger} =
   \frac{1}{\sqrt{2}}
   \left( \frac{\hat{q}}{b}
     - i \frac{\hat{p}}{c}
   \right), \qquad
z_0 =
   \frac{1}{\sqrt{2}}
   \left(
     \frac{q_0}{b}
     + i \frac{p_0}{c}
   \right) \;.
\end{equation}
Here $| 0 \rangle$ is the ground state of a harmonic oscillator of
frequency $\omega=\hbar/m b^2$, $\hat{a}^{\dagger}$ is the
creation operator and $q_0$, $p_0$ are the mean values of the
position $\hat{q}$ and momentum $\hat{p}$ operators, respectively.
The widths in position $b$ and momentum $c$ satisfy $b c = \hbar$
and $z_0$ is complex. The semiclassical approximation for the
coherent state propagator ${K}(z_f^*, z_0, T) \equiv \langle z_f |
e^{-i\hat{H}T/\hbar} | z_0 \rangle$ is given by
\cite{Klau78,Weis82b,Bar01}
\begin{eqnarray}
\label{eq1} {K} \left( z_f^*,z_0,T \right) \approx
   \sum_{\mathrm{traj.}}
   \sqrt{ \frac{1}{\left|\mathrm{M}_{vv} \right|}} \,
   \exp{ \left\{ \frac{i}{\hbar} \, F \right\}},
\end{eqnarray}
where $\mathrm{M}_{vv}$ and $F$ depend on (generally complex)
classical trajectories. These trajectories are best represented in
terms of the variables $u$ and $v$, instead of the canonical
variables $q$ and $p$, defined by
\begin{eqnarray}
\label{eq2} u =
    \frac{1}{\sqrt{2}}
    \left(\frac{q}{b} + i \frac{p}{c} \right)
    \qquad \mathrm{and} \qquad v =
    \frac{1}{\sqrt{2}}
    \left( \frac{q}{b} - i \frac{p}{c} \right) .
\end{eqnarray}
The sum in Eq.~(\ref{eq1}) runs over all trajectories governed by
the complex Hamiltonian $\tilde{H}(u,v) \equiv \langle v | \hat{H}
| u \rangle$ and satisfying the boundary conditions $u(0) \equiv
u' = z_0$ and $v(T) \equiv v'' = z_f^*$. Notice that $q$ and $p$
are complex variables, while the propagator labels, $q_0$, $p_0$
for the initial state and $q_f$, $p_f$ for the final one, are
real. In Eq. (\ref{eq1}), $F$ is given by
\begin{eqnarray}
F( v'',  u', T) =
   \mathcal{S}(v'', u', T) +
   \mathcal{G}( v'', u', T) -
   \frac{\hbar}{2} \sigma ,
\label{eq3}
\end{eqnarray}
where $\mathcal{S}$, the complex action of the trajectory, and
$\mathcal{G}$ are given by
{\small
\begin{displaymath}
\mathcal{S}( v'', u', T) =
   \int_{0}^{T}
    \left[
      \frac{i \hbar}{2}
      \left(\dot{u} \, v - u \, \dot{v} \right) -
      \tilde{H}
   \right] dt
   - \frac{i \hbar}{2} \left[ u'' v'' + u'v' \right],
\end{displaymath}
\begin{displaymath}
\mathcal{G}( v'', u', T) =
   \frac{1}{2} \int_{0}^{T}
      \frac{\partial^{2}\tilde{H}}{\partial u \; \partial v}
    \, dt \; .
\end{displaymath}
}
Finally $\mathrm{M}_{vv}$, and its phase $\sigma$, is an element
of the tangent matrix defined by
\begin{eqnarray}
\left(
   \begin{array}{c}
   \delta u''\\
   \delta v''\\
   \end{array}
\right) = \left(
   \begin{array}{cc}
   \mathrm{M}_{uu}   & \mathrm{M}_{uv}   \\
   \mathrm{M}_{vu}   & \mathrm{M}_{vv}   \\
   \end{array}
\right) \left(
   \begin{array}{c}
   \delta u'\\
   \delta v'\\
   \end{array}
\right) \, , \label{eq5}
\end{eqnarray}
where $\delta u$ and $\delta v$ are small displacements around the
complex trajectory. We use a single (double) prime to indicate
initial time $t=0$ (final time $t=T$). The elements of the tangent
matrix can also be written in terms of second derivatives of the
action (see Ref. \cite{Bar01}). Note that Eq.(\ref{eq3}) differs
from the formula given in \cite{Bar01} because we are using
non-normalized coherent states.

Phase space caustics occur when $\mathrm{M}_{vv} =0$, causing the
semiclassical propagator to diverge. Close to these points the
semiclassical formula provides only a poor approximation to the
quantum propagator. For a discussion of the mechanisms that lead
to caustics in systems with one degree of freedom see
\cite{Klau95,Agu05,parisio2}.

Caustics in the semiclassical propagator in a given representation
can usually be circumvented by applying the Maslov method. This
requires the calculation of the semiclassical propagator in the
respective conjugate representation, followed by the
transformation back to the original one, with this last step
performed with an approximation better than quadratic. For the
case of coherent states, although there is no natural dual
representation to be used with $K(v'',u',T)$, the complex action
$\mathcal{S}(v'',u',T)$ satisfies the relations
\begin{eqnarray}
u(T)  =
   \frac{i}{\hbar}
   \frac{\partial \mathcal S}{\partial v''}
\qquad \mathrm{and} \qquad v(0) =
   \frac{i}{\hbar}
   \frac{\partial \mathcal S}{\partial u'} \, ,
\label{ull}
\end{eqnarray}
which suggests a Legendre transformation $\mathcal{S} \rightarrow
\tilde{\mathcal{S}}$, by the change of variables $v''\rightarrow
(i/ \hbar)(\partial \mathcal S /\partial v'')$. The transformed
function $\tilde{\mathcal{S}}$ depends on $u'$ and $u''$, instead
of $u'$ and $v''$,
\begin{eqnarray}
\tilde{\mathcal S} ( u'', u',T)
  = \mathcal S(v'', u',T ) + i \hbar u'' v'' \, ,
\label{Tlegen}
\end{eqnarray}
and satisfies the relations
\begin{eqnarray}
v'' =
   -\frac{i}{\hbar}
   \frac{\partial \tilde{\mathcal S}}{\partial u''}
   \qquad \mathrm{and} \qquad
v' =
   \frac{i}{\hbar}
   \frac{\partial \tilde{\mathcal S}}{\partial u'}  \, .
\label{stilpartial}
\end{eqnarray}
These properties, on the other hand, suggest the following
definition for the dual representation of the semiclassical
propagator:
\begin{eqnarray}
\tilde{K} (u'', u', T)  =
   \frac{1}{\sqrt{2 \pi i}} \int_C{
      K(v'', u', T ) e^{-u'' v''}
      \mathrm{d} v''
   }  \label{ktil}
\end{eqnarray}
where the path $C$ will be specified below.

In the semiclassical limit this integral can be solved by the
steepest descent method \cite{bleistein} and an explicit
expression for $\tilde{K}$ can be obtained. Inserting
Eq.(\ref{eq1}) into (\ref{ktil}) we find the saddle point
condition
\begin{equation}
\frac{\partial}{\partial v''} \left[ \mathcal{S} + i\hbar u''
v''\right] =0 \qquad
  \mbox{or} \qquad u'' =
   \frac{i}{\hbar}\frac{\partial \mathcal{S}}{\partial v''},
\label{crit}
\end{equation}
where we have considered that $\mathcal{G}$ varies slowly in
comparison with $\mathcal{S}$ (see Ref. \cite{Bar01}).
Eq.(\ref{crit}) says that the stationary trajectory satisfies
$u(0)=u'$ and $u(T)=u''$, i.e., the saddle point value $v_c''$ of
the integration variable is equal to $v(T)$ of a trajectory
satisfying these boundary conditions. This imposes that the
integration path $C$ must coincide with (or be deformable into) a
steepest descent path through $v_c''$. Expanding the exponent up
to second order around this trajectory and performing the Gaussian
integral we obtain
\begin{eqnarray}
\tilde{K}(u'', u', T ) \!=\!
   \sqrt{\frac{i}{| \mathrm M_{u v}|}} \;
   e^{
      \frac{i}{\hbar} \tilde{\mathcal S}
      (u'', u', T )
      + \frac{i}{\hbar} \tilde{\mathcal G}
      (u'', u',T ) - \frac{i}{2}
      \tilde{\sigma}}. \label{eq9}
\end{eqnarray}
We emphasize that $\tilde{K}$ depends on classical trajectories
satisfying $u'=u(0)$ and $u'' =u(T)$. $\mathrm M _{u v}$ is given
by Eq. (\ref{eq5}), $\tilde{\sigma}$ is its phase,
$\tilde{\mathcal G}(u'', u',T) $ is the function $\mathcal{G}$
calculated at the new trajectory, and $\tilde{\mathcal S}(u'',
u',T )$ is given by Eq. (\ref{Tlegen}). The PSC affecting
$\tilde{K}$ correspond to trajectories for which $M_{uv}=0$, which
generally do not coincide with the PSC of $K$.

The inverse transformation of Eq.(\ref{ktil}) is given by
\begin{eqnarray}
K(v'', u', T )  =
   \frac{1}{\sqrt{2 \pi i}}
   \int_{\tilde{C}}{
      \tilde{K}(u'', u', T ) e^{u'' v'' }
      \mathrm{d} u''
   }\;.
\label{eq8}
\end{eqnarray}
Replacing Eq.(\ref{eq9}) into (\ref{eq8}) and doing the integral
again by the steepest descent approximation up to second order
gives back the original propagator of Eq.(\ref{eq1}).

The pairs of equations (\ref{eq1})-(\ref{eq9}) and
(\ref{ktil})-(\ref{eq8}) look very much like the corresponding
transformation for the propagators in coordinates and momenta,
$K(x_f,x_0,T)$ and $K(p_f,x_0,T)$. In that case the classical
trajectory goes from $x_0$ to $x_f$ in one case and from $x_0$ to
$p_f$ in the other. However, although the coherent states define a
true quantum mechanical representation \cite{bargmann}, and the
propagator $K(z_f^*,z_0,T)$ corresponds to the matrix element
$\langle z_f|e^{-i\hat{H}T/\hbar}|z_0\rangle$, there is no
representation such that $\tilde{K}(z_f,z_0,T)$ also corresponds
to a similar matrix element. $\tilde{K}$ would involve two kets,
$|z_0\rangle$ and $|z_f\rangle$ instead of a ket $|z_0\rangle$ and
a bra $\langle z_f|$.

Therefore, since $\tilde{K}$ is {\it not} a matrix element of the
evolution operator we must formalize its quantum mechanical
definition so that the previous transformations make precise
sense. This is done as follows: given a ket $|f\rangle$ and its
Bargmann representation $f(z^*) = \langle z|f \rangle$
\cite{bargmann}, for each coherent state ket $|w\rangle$ we define
the application
\begin{equation}
\tilde{f}(w) =\! \frac{1}{\sqrt{2 \pi i}}\int_\gamma \frac{\langle
z|f \rangle}{\langle z|w \rangle} dz^* = \!\frac{1}{\sqrt{2 \pi
i}}\int_\gamma f(z^*) \, e^{-z^* w} dz^* . \label{ftil}
\end{equation}
The path $\gamma$ must be chosen in such a manner that the
integral becomes a Laplace transform. This definition is
suggestive of the need to, so to speak, ``cancel the bra $\langle
z|$ and replace it by a ket $|w\rangle$''. At the same time it
provides just the right Legendre transform we need in the
semiclassical limit when $|f\rangle = e^{-iHT/\hbar} |z_0\rangle$.
The inverse transformation is defined as
\begin{equation}
f (z^*) =\! \frac{1}{\sqrt{2\pi i}} \int_{\gamma'} \tilde{f}(w)
\langle z|w \rangle dw =\! \frac{1}{\sqrt{2\pi i}} \int_{\gamma'}
\tilde{f}(w) \, e^{z^* w} dw \label{ftilinv}
\end{equation}
with $\gamma'$ chosen so that the integral is a Mellin transform.

To illustrate the transformation we apply it to the harmonic
oscillator. Let $|f\rangle$ be an eigenstate $|m\rangle$ of the
Hamiltonian operator. Then $\langle z | m \rangle \equiv \phi_m
(z^*) = (z^*)^m/\sqrt{m!}$ ~and
\begin{equation}
\tilde{\phi}_m (w) = \frac{1}{\sqrt{2\pi i \, m!}} \int_\gamma
                     (z^*)^m e^{-z^* w} dz^*.
\label{ohint}
\end{equation}
Writing $w=|w|e^{i\theta}$ and  $z^* = r e^{-i\alpha}$, the path
$\gamma$ is defined by $\alpha=\theta$ with $r$ varying from $0$
to $\infty$. This produces
\begin{equation}
\tilde{\phi}_m (w) = \frac{1}{\sqrt{2\pi i}}
\frac{\sqrt{m!}}{w^{m+1}}. \label{ohint2}
\end{equation}
The inverse transformation is given by
\begin{equation}
\breve{\phi}_m(z^*) = \frac{\sqrt{m!}}{2\pi i}
              \int_{\gamma'}
              \frac{e^{z^* w}}{w^{m+1}} dw .
\end{equation}
Writing $z=|z|e^{i\phi}$ and choosing $\gamma'$ so that $w =
(-\alpha+it)e^{i\phi}$, with $\alpha >0$ fixed and $t$ varying
from $-\infty$ to $+\infty$, we can solve the integral by the
method of residues and we find exactly $\breve{\phi}_m(z^*) =
\phi_m(z^*)$. For the propagator we set $|f\rangle =
e^{-iHT/\hbar} |z_0\rangle$ and obtain
\begin{equation}
K(z_f^*,z_0,T)\equiv \langle z_f | e^{-i\hat{H}T/\hbar} |
z_0\rangle =
    e^{z_0 z_f^* e^{-i\omega T}-i \omega T/2}
\label{exact}
\end{equation}
and
\begin{equation}
\tilde{K}(w,z_0,T) = \frac{1}{\sqrt{2\pi i}} \, \frac{e^{-i \omega
T/2}}{w-z_0 e^{-i\omega T}}\;. \label{ktilosc}
\end{equation}

Equations (\ref{ftil}) and (\ref{ftilinv}) show that the
semiclassical propagator $\tilde{K}$ given by Eq.(\ref{eq9}) is
the semiclassical approximation of a true quantum mechanical
function, namely Eq.(\ref{ftil}) with $f(z^*)=K(z^*,z_0,T)$. This
function, although not a matrix element in a mixed representation
like $K(p_f,x_0,T) = \langle p_f|e^{-iHT/\hbar}|x_0\rangle$, is
well defined provided the integral over $\gamma$ converges. The
application defined by Eqs.(\ref{ftil}) and (\ref{ftilinv}) can be
thought of as conjugate to the Bargmann representation, and they
provide the tools to the application of the Maslov method to the
coherent state propagator.

The connection between the propagator, Eq.(\ref{eq1}), and its
conjugate function, Eq.(\ref{eq9}), via steepest descent
approximation with quadratic expansion of the exponent works only
in the regions where both ${\mathrm M _{u v}}$ and ${\mathrm M _{v
v}}$ are non-zero. Close to caustics, where two stationary
trajectories coalesce and ${\mathrm M _{v v}}=0$, $\tilde{K}$ is
still well defined and $K$ can be obtained by doing the inverse
transform (\ref{eq8}) using a uniform approximation \cite{berry}.
The basic idea is to map the function in the exponent of the
integrand into an auxiliary cubic function of a new variable $X$.
The new function is chosen in such a way that its stationary
points coincide with those of the original function. Inserting Eq.
(\ref{eq9}) into Eq. (\ref{eq8}) we define the new integration
variable $X = X(u'')$ by
\begin{eqnarray}
\frac{1}{\hbar} [\tilde{\mathcal S}+\tilde{R}] -i u''v'' \equiv A
- BX + X^3/3 \label{expx}
\end{eqnarray}
where $\tilde{R}=\tilde{\mathcal G} + (i\ln{|M_{uv}|}
-\tilde{\sigma})\hbar/2$ contains the slowly varying terms and $A$
and $B$ are functions of $u'$, $v''$ and $T$. Differentiating both
sides with respect to $X$ and discarding the variation of
$\tilde{R}$ yields
\begin{eqnarray}
\left[\frac{1}{\hbar} \frac{\partial \tilde{\mathcal S}}{\partial
u''} - iv''\right] \frac{\partial u''}{\partial X}\! = i[v(T)-v'']
\frac{\partial u''}{\partial X} = -B + X^2 \!.\label{expdx}
\end{eqnarray}
The stationary condition $v(T)=v''$ has generally two solutions,
$u_+''$ and $u_-''$, in the vicinity of a caustic. These two
stationary points, that coalesce at the caustic, are mapped into
$X_{\pm}=\pm B^{1/2}$, while the caustic itself corresponds to
$X=0$. Substituting $X=X_{\pm}$ in (\ref{expx}) and solving for
$A$ and $B$ we find
\begin{eqnarray}
A = \frac{1}{2\hbar}(S_++\tilde{R}_++S_-+\tilde{R}_-) \;, \nonumber \\
B=\left[-\frac{3}{4\hbar}(S_++\tilde{R}_+-S_--\tilde{R}_-)
\right]^{2/3} \label{ab}
\end{eqnarray}
where $S_{\pm}$ is the action $S=\tilde{S} - i\hbar u''v''$
calculated at the stationary trajectories defined by $u_{\pm}''$.

The change of variables from $u''$ to $X$ also produces a Jacobian
$f(X)\equiv \partial u''/\partial X$. Since the $X$ intervals that
contribute significantly to integral are those close to the
stationary points, we need to specify the Jacobian only in these
regions. Writing $f(X) = C + G (X-B^{1/2}) + H (X+B^{1/2})$ and
defining $f_{\pm} = f(X_{\pm})$ we find
\begin{eqnarray}
f(X)=(f_++f_-)/2 + X(f_+-f_-)/2B^{1/2} \,. \label{fff}
\end{eqnarray}
Differentiating (\ref{expdx}) with respect to $X$ once again and
calculating at $X_{\pm}$ we obtain
\begin{eqnarray}
f_{\pm} = \left(\pm\frac{2 M^{\pm}_{uv} B^{1/2}}{i M^{\pm}_{vv}}
\right)^{1/2}.
\end{eqnarray}
Putting everything together we find the following uniform
approximation for the propagator
\begin{eqnarray}
K(u',v'',T) = \frac{1}{\sqrt{2\pi}}\int f(X) \; e^{i(A-BX+X^3/3)}
d X\,. \label{final}
\end{eqnarray}

Far from the caustic, where the contribution of each stationary
trajectory can be computed separately, $f(X)$ reduces to $f_{\pm}$
and the integral can be written in terms of an Airy function. In
this case the Airy function can be evaluated by the method of
steepest descent, with the integration path chosen according to
the phase of its argument $z=-B$. Therefore the argument of $z$
automatically indicates whether the two stationary points of the
exponent contribute to the propagator or if only one of them do.
Expanding the exponent to second order about the contributing
points ($X_+$ or $X_-$ or both, depending on the arg$(z)$), and
doing the resulting Gaussian integral recovers the quadratic
approximation Eq.(\ref{eq1}). As an illustration we consider the
Hamiltonian $\hat{H}=(a^{\dagger}a+1/2)^2$. Fig.1 shows the square
modulus of the diagonal propagator $\langle z| e^{-i\hat{H}T} |z
\rangle$ ($b=c=\hbar=1$) as a function of $T$ for
$z=1/(2\sqrt{2})$. The dotted line displays the bare semiclassical
result, showing a large increase for $T\gtrsim 2.0$, due to a
nearby caustic. The exact result is the full line and the uniform
approximation is shown by the dashed line, which is indeed
uniformly good at all times. Detailed numerical applications will
be published elsewhere.

\begin{figure}
\includegraphics[width=6cm,angle=-90]{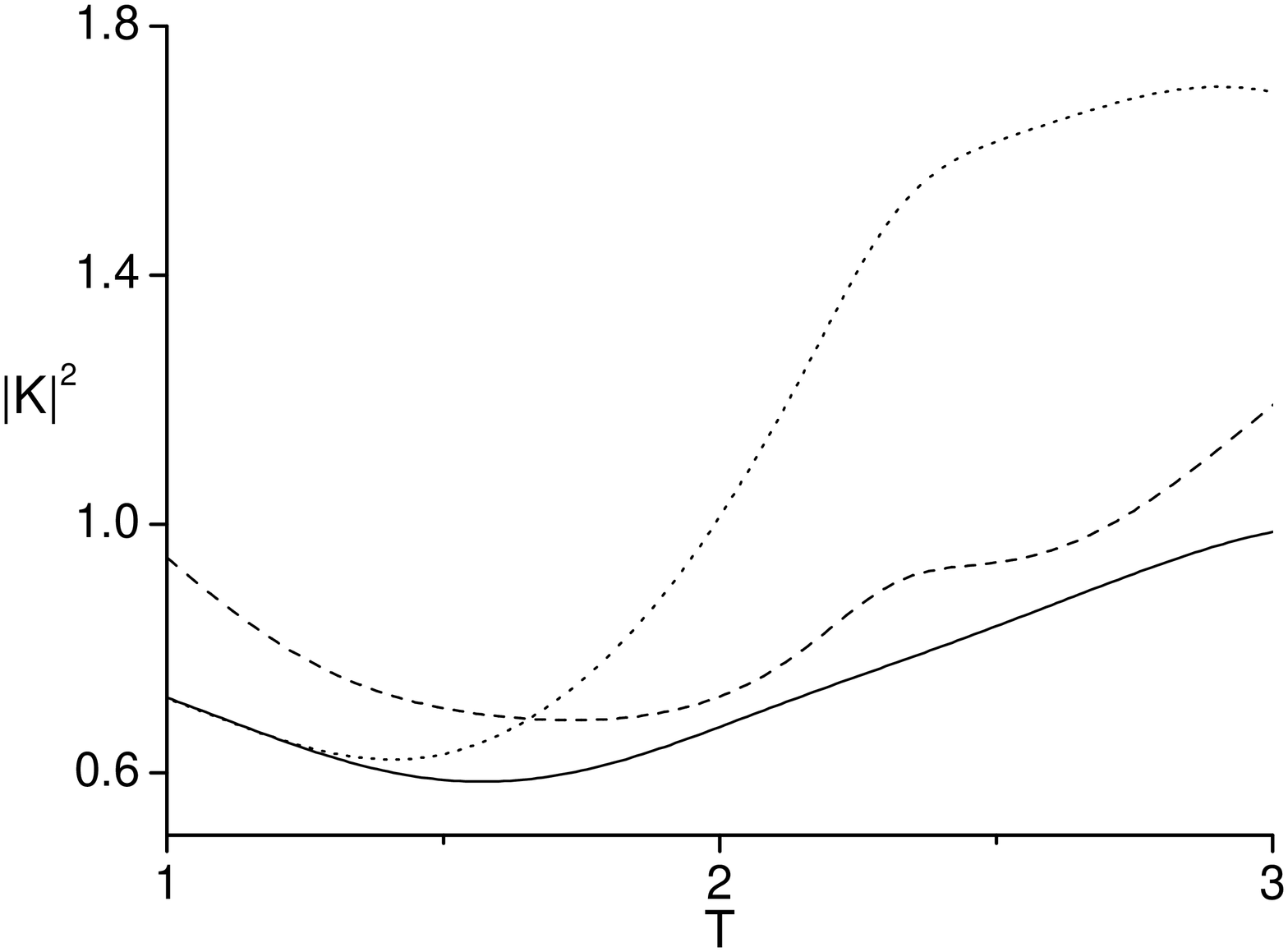}
\caption{Square modulus of diagonal propagator for fixed $z$. The
lines correspond to the exact result (full), bare semiclassical
(dotted) and uniform (dashed).}
\end{figure}

Eq.(\ref{final}) and the definition of the conjugate application
and its inverse, Eqs.(\ref{ftil}) and (\ref{ftilinv}), constitute
the main results of this paper. Although the idea of a conjugate
application to the Bargmann representation is used here just as a
tool to derive the above uniform approximation, it may be useful
in other situations. For instance, a transitional approximation,
valid only close to the caustics, can also be derived. In
addition, the Fourier frequencies of the transformed propagator
are the eigenvalues of the Hamiltonian, and it might be simpler to
extract those eigenvalues from the transformed propagator than
from the Bargmann propagator. This is certainly the case for the
Harmonic oscillator, since the time dependence of
$\tilde{K}(w,w,T)$, see Eq.(\ref{ktilosc}), is trivial.

Acknowledgments: This work was partly supported by FAPESP and
CNPq.

\end{document}